\documentclass[11pt]{article}

\usepackage{epsfig}
\usepackage{subfig}
\usepackage{graphicx}
\usepackage{epstopdf}
\usepackage{amsfonts}
\usepackage{amssymb}
\usepackage{amsthm}
\usepackage{amsmath}
\usepackage{multirow}
\usepackage{color}
\usepackage{cite}

\def\beq{\begin{equation}}
\def\eeq{\end{equation}}
\def\bea{\begin{eqnarray}}
\def\eea{\end{eqnarray}}
\newcommand{\beqs}{\begin{subequations}}
\newcommand{\eeqs}{\end{subequations}}

\newcommand{\cref}[1]{Ref.~\cite{#1}}

\newcommand{\hh}{{\ensuremath{I{\kern-2.6pt h}}}}
\newcommand{\bhh}{{\ensuremath{\bar{I{\kern-2.6pt h}}}}}

\usepackage[colorlinks=true,allcolors=blue]{hyperref}

\begin{document}

\begin{titlepage}
	

\begin{center}
{\Large {\bf Gravitational Waves from Quasi-stable Strings}
}
\\[12mm]
George Lazarides,$^{1}$~
Rinku Maji,$^{2}$
Qaisar Shafi$^{3}$~
\end{center}
\vspace*{0.50cm}
\centerline{$^{1}$ \it
School of Electrical and
Computer Engineering, Faculty of Engineering,
}

\centerline{\it
Aristotle University
of Thessaloniki, Thessaloniki 54124, Greece}
\vspace*{0.2cm}
	\centerline{$^{2}$ \it
		Theoretical Physics Division, Physical Research Laboratory,}
		\centerline{\it  Navarangpura, Ahmedabad 380009, India}
	\vspace*{0.2cm}
	\centerline{$^{3}$ \it
		Bartol Research Institute, Department of Physics and 
		Astronomy,}
	\centerline{\it
		 University of Delaware, Newark, DE 19716, USA}
	\vspace*{1.20cm}
\begin{abstract}
 We estimate the stochastic gravitational wave spectrum emitted from a network of cosmic strings in which the latter are effectively stable against breaking by monopole pair creation. The monopoles are produced at a higher scale from an earlier symmetry breaking and experience significant inflation before reentering the horizon. This gives rise to monopole-antimonopole pairs connected by string segments and the string loop formation essentially ceases. As a consequence, the lower frequency portion of the gravitational wave spectrum is suppressed relative to the no-inflation case with stable strings, which evades the stringent PPTA bound on the dimensionless string tension $G\mu$. We display the modified spectrum, accessible in the ongoing and future experiments, for $G\mu$ values in the range $10^{-10} - 10^{-15}$. We show how this `quasi-stable' string network is realized in realistic grand unified theories.

\end{abstract}

\end{titlepage}
\section{Introduction}
Topologically stable strings with varying mass scales can appear in
Grand Unified Theories (GUTs) and a particularly compelling example
is provided by $SO(10)$ broken to the Standard Model (SM) by only using
Higgs fields in the tensor representations of the gauge group \cite{Kibble:1982ae}. Independent of the SO(10) symmetry breaking pattern a gauged $Z_2$
symmetry in this case remains unbroken, which gives rise to topologically stable $Z_2$ cosmic strings. The string tension depends on the appropriate symmetry breaking scale which can vary between $M_{\rm GUT}$ ($\sim 10^{16}$ GeV) and scales as low as a TeV or so. It is worth repeating here that this $Z_2$ symmetry, in a supersymmetric $SO(10)$ setting, is precisely matter parity and thereby yields a compelling dark matter candidate (lightest supersymmetric particle). 

Topologically stable cosmic strings with mass per unit length $\mu\sim
M^2$, where $M$ denotes the relevant symmetry breaking scale, are severely constrained by a variety of experimental observations related to cosmic microwave background radiation \cite{Charnock:2016nzm,Lizarraga:2016onn, Torki:2021fvi} and pulsar timing arrays \cite{Blanco-Pillado:2017rnf, Ringeval:2017eww}. In particular, for strings associated with the breaking of a local gauge symmetry, the dimensionless string tension parameter $G\mu \lesssim 4.6 \times 10^{-11}$ \cite{Lazarides:2021uxv}, where $G$ denotes Newton’s constant. In other words, topologically stable cosmic strings associated with a mass scale larger than $4.7 \times 10^{13}$ GeV or so emit stochastic gravitational radiation \cite{Hindmarsh:1994re, Vilenkin:2000jqa, Olmez:2010bi, Blanco-Pillado:2017oxo,Auclair:2019wcv} that should have been seen with the current detectors.

One way to evade the above constraint on $G\mu$ relies on a metastable cosmic string network, which allows one to manipulate the gravitational wave spectrum and make it compatible with the observations despite the larger symmetry breaking scales \cite{Buchmuller:2019gfy,Buchmuller:2020lbh,Buchmuller:2021dtt,Buchmuller:2021mbb,Masoud:2021prr,Dunsky:2021tih,Ahmed:2022rwy,Afzal:2022vjx}. The string  metastability arises from the fact that it can break through the formation of monopole-antimonopole pairs. The monopole scale is not far above the string scale in order for this process to occur at a reasonable rate.

In another approach \cite{Guedes:2018afo, Cui:2019kkd,Chakrabortty:2020otp, Lazarides:2021uxv} the cosmic strings undergo partial inflation before reentering the horizon such that the modified spectrum becomes compatible with the observations for $G\mu > 4.6 \times 10^{-11}$.

In this paper we estimate the stochastic gravitational wave background emitted by a quasi-stable cosmic string network in the presence of monopoles. The scales associated with the monopoles and strings can be separated in magnitude such that quantum tunneling by monopole-antimonopole pair creation is ineffective and the strings are practically stable.
The monopoles are created at a phase transition prior to the strings and are partially inflated. At a subsequent transition, a string network is generated connecting the monopoles to antimonopoles.
(The strings may also connect two different types of monopoles if more types of monopoles are produced.)  The strings can also be partially inflated. After reentering the horizon they are actually random walks with a step equal to the horizon scale at any given time. The monopoles and antimonopoles eventually also reenter the  horizon connected by string segments with length of order the horizon size. Gravitational waves are generated by decaying string loops which were created before the monopoles reentered the horizon as well as decaying string segments after that. We show how $G \mu$ values up to about $10^{-10}$ are compatible with the current observations, and how this scenario can be tested in future experiments for a wide range of the dimensionless string parameter. An 
$SO(10)$ and a trinification model where this scenario can be realized are briefly discussed.

\section{Cosmology of Quasi-stable strings}

We assume here a non-supersymmetric GUT model with at least two intermediate breaking scales. (By intermediate scale we mean a symmetry breaking scale that does not exceed the Hubble parameter during inflation, namely $H_{\rm inf}\lesssim 10^{14}$ GeV \cite{Planck:2018jri}). The first scale $M_I$ is associated with magnetic monopole production, while the second one $M_{II}$ leads to cosmic strings which connect monopoles and antimonopoles. Clearly, the strings are not topologically stable since they can break by monopole pair creation. However, if the monopole mass is adequately higher than the string scale, this quantum tunneling  process  is utterly suppressed and the strings are stable for all practical purposes (see  discussion below for why we call such strings 
`quasi-stable' instead of the widely used `metastable'). The monopoles are also not topologically stable since they can be pulled together by the connecting string segments and annihilate. They are produced during the first intermediate scale phase transition and are
partially inflated. During the second intermediate transition a network of quasi-stable cosmic strings is generated. This may happen either during inflation with the strings also being partially inflated, or after the end of inflation. In any case, the strings will enter the post-inflationary horizon or be generated for the first time at a relatively early cosmic time $t_F$. At subsequent times, the strings form random walks with a step of the order of the horizon scale and inter-commute, thereby generating loops which decay into gravitational waves. At a later time $t_M$, the monopoles enter the horizon forming monopole-antimonopole pairs connected by string segments with length of order the horizon size. These segments also decay predominantly into gravitational waves contributing to their overall spectrum provided that the monopoles do not carry unconfined fluxes. 
Since in this scenario the long string loops or segments are absent, the gravitational wave spectrum at lower frequencies is reduced. Consequently, the predicted spectrum can be compatible with the pulsar timing array bounds for a wider range of $G\mu\simeq (1/8)(M_{II}/m_{\rm Pl})^2$ values and at the same time be accessible at the ongoing and planned experiments ($m_{\rm Pl}$ is the reduced Planck mass).

The quasi-stable cosmic strings can break, in principle, via the Schwinger production of monopole-antimonopole pairs with a rate per unit length given by \cite{Leblond:2009fq,Monin:2008mp,Monin:2009ch}
\begin{equation}
\Gamma_d=\frac{\mu}{2\pi}\exp{\left(-\pi m_M^2/\mu\right)} \, ,
\end{equation}
where $m_M\simeq 4\pi M_{I}/g^2$ is the monopole mass ($g$ is the gauge coupling). The decay width of a string loop or segment of size $l$ to gravitational waves is of order $(\mathcal{C} G\mu)l^{-1}$, where $\mathcal{C}$ is a numerical factor of $\mathcal{O}(10^2)$ \cite{Vachaspati:1984gt,Martin:1996cp}. For string loops $\mathcal{C}=\Gamma\sim 50$ \cite{Vilenkin:2000jqa}. Quasi-stability of the strings requires that this decay width is much bigger than $\Gamma_{d}l$. In our case where loops and segments larger than of order $t_M=10^{10}~{\rm sec}$ are not considered, this requirement is fulfilled provided that the ratio of the monopole mass to the string scale $r_{ms}=(m_M^2/\mu)^{1/2}$ is larger than about 8.7.

In Ref.~\cite{Buchmuller:2020lbh}, the authors considered metastable strings with the above ratio in the range $7.9-8.15$. These strings at some point decay by monopole pair creation and thus large loops are absent. This reduces the gravitational wave spectrum at lower frequencies and allows higher values of $G\mu$ leading to a large stochastic gravitational wave signal in the frequency band of the ongoing and planned experiments. They indicate, consistent with our findings, that for $r_{ms}\gtrsim 9$ the strings become practically stable. Our approach is different since the strings are practically stable but large loops are avoided due to the reentrance of the partially inflated monopoles into the horizon. This latter scenario is realized in realistic extensions of the SM that we briefly discuss.
                   
\section{Stochastic Gravitational Wave Background}
We consider a scenario where the monopoles are partially inflated which is followed by the appearance in the next transition of cosmic strings. The strings connect monopoles to antimonopoles and perform random walks with a step of the order of the horizon scale at their formation. 
They are subsequently also partially inflated together with the monopole. The strings reenter the horizon at a relatively early time $t_F$ after inflation. They form random walks with a step of the order of the horizon scale at subsequent times. Therefore, they inter-commute and form loops of size 
$\alpha t_i$ ($\alpha\sim 0.1$) at any subsequent time $t_i$, which decay via the radiation of gravitational waves. The length of the decaying string loop at any time $t$ is given by
\begin{align}\label{eq:loop-length}
l(t)=\alpha t_i - \Gamma G\mu (t-t_i),
\end{align}
where $\Gamma \sim 50$ is a numerical factor.

As the monopoles reenter the horizon at time $t_M$ during the radiation dominated Universe, the loop formation stops and the monopole-antimonopole pairs connected by string segments ($MS\bar{M}$) decay by emitting gravitational waves. Large string loops are consequently absent in this case, and thus the lower frequency part of the gravitational wave spectrum from the loops is reduced.
The loop distribution $n(l,t)$, which is the number of loops per unit volume and per unit loop size at loop length $l$ and time $t$ during radiation domination, is expressed as \cite{Blanco-Pillado:2013qja,Blanco-Pillado:2017oxo} (also see the supplementary material of Ref.~\cite{LIGOScientific:2021nrg})
\begin{align}\label{eq:n-loop-rad}
n_r(l,t<t_M) = \frac{0.18 \ \Theta(0.1 t-l)}{t^{3/2}(l+\Gamma G\mu t)^{5/2}} \, , \quad \mathrm{for} \ t<t_M .
\end{align} 
After $t_M$ we will be left with string segments of initial length $\sim 2t_M$ which connect monopole-antimonopole pairs. We have assumed that the strings formed initially are predominantly open connecting monopoles to antimonopoles and there are no unconfined fluxes carried by the monopoles. There will be some loops formed from the $MS\bar{M}$ structures. The intercommutation probability of the subhorizon segments is small in comparison with that of the superhorizon strings before $t_M$. Therefore, we expect that the contribution from such loops will be suppressed. Moreover, these loops are soon becoming much smaller than the horizon size since the $MS\bar{M}$ structures after $t_M$ behave like a gas of pressureless decaying particles and their absolute size can only decrease.  They may contribute moderately to the high frequency spectrum from the $MS\bar{M}$ structures, which, as we will see, is anyway suppressed. The distribution of loops which are remnants of those formed before $t_M$ is given by (see Appendix for derivation),
\begin{align}\label{eq:n-loop-rad_after_tM}
n_r(l,t>t_M) = \frac{0.18 \ \Theta(0.1 t_M-l-\Gamma G\mu(t-t_M))}{t^{3/2}(l+\Gamma G\mu t)^{5/2}} \quad \mathrm{for} \ t_{eq}>t>t_M \, ,
\end{align}
or
\begin{align}\label{eq:n-loop-radmat}
n_{rm}(l,t>t_M) = \frac{0.18 t_{eq}^{1/2}\Theta(0.1 t_M-l-\Gamma G\mu(t-t_M))}{t^2(l+\Gamma G\mu t)^{5/2}} \quad \mathrm{for} \ t > t_{eq} \, , 
\end{align}
in the radiation or matter dominated era that follows $t_M$. 
Here $\Theta$ is the Heaviside function and $t_{eq}$ denotes the equidensity time when the radiation and matter densities coincide. 
  
The stochastic gravitational wave background receives contributions from the oscillating string loops with their distribution given in Eq.~(\ref{eq:n-loop-rad}) before $t_M$. After $t_M$, the contributions mainly come from the decaying string loops formed before $t_M$, as given in 
Eqs.~(\ref{eq:n-loop-rad_after_tM}) and (\ref{eq:n-loop-radmat}), and from the oscillating $MS\bar{M}$ structures. 

 The gravitational wave spectra are computed by superimposing the unresolved gravity wave bursts \cite{Olmez:2010bi, Leblond:2009fq, Auclair:2019wcv} and this is a direct method. In alternative methods, the gravitational wave spectra are expressed as a sum of the contributions from an infinite number of normal modes 
\cite{Vachaspati:1984gt, Blanco-Pillado:2013qja,Blanco-Pillado:2017oxo,Cui:2018rwi}. On technical ground, however, we have to take a finite number of modes to compute the spectra. We can achieve the required accuracy by taking the sum of around $k\sim 10^5$ modes \cite{Cui:2019kkd}. For higher frequencies we need to take a much bigger number of modes depending on $t_F$ (see supplementary material of Ref.~\cite{Cui:2019kkd} and Ref.~\cite{Blasi:2020wpy} for more details.) As argued in Ref.~\cite{Cui:2019kkd}, the spectra match with the burst method if a sufficiently large number of normal modes is taken into account.
\subsection{Gravitational Waves from String Loops}
\label{sec:gws-loops}
Assuming cusp domination, the emitted gravitational wave background at frequency $f$ is given by \cite{Olmez:2010bi, Auclair:2019wcv, Cui:2019kkd, LIGOScientific:2021nrg}
\begin{align}\label{eq:GWs-Omega-cusps}
\Omega^{\rm loop}_{GW}(f) = \frac{4\pi^2}{3H_0^2}f^3\int_{z_{\rm min}}^{z(t_F)} dz \, \Theta(t_i-t_F) \int dl \, h^2(f,l,z)\frac{d^2R}{dz \, dl} \ ,
\end{align}
where the burst rate per unit space-time volume is
\begin{align}
\frac{d^2R}{dz \, dl} = N_c H_0^{-3}\phi_V(z) \frac{2n(l,t(z))}{l(1+z)}\left( \frac{\theta_m(f,l,z)}{2}\right)^2\Theta(1-\theta_m),
\end{align}
with the beam opening angle
\begin{align}
\theta_m(f,l,z) = \left[\frac{\sqrt{3}}{4}(1+z)fl\right]^{-1/3},
\end{align}
and $n(l,t)$ is given by Eqs.~(\ref{eq:n-loop-rad}), (\ref{eq:n-loop-rad_after_tM}), and (\ref{eq:n-loop-radmat}) at the appropriate cosmic era. The quantity $t_i$ denotes the time of loop formation and is found from Eq.~(\ref{eq:loop-length}) for any redshift $z(t)$, and $H_0$ is the present value of the Hubble parameter. The theta function in Eq.~(\ref{eq:GWs-Omega-cusps}) ensures that we only consider the loops formed after the horizon reentry of the string network, which is assumed to lie in the radiation dominated era (see Sec.~\ref{sec:results} for details). We have taken $N_c=2.13$ as in Ref.~\cite{Cui:2019kkd}. The lower limit $z_{\rm min}$ in the integral in Eq.~(\ref{eq:GWs-Omega-cusps}) separates the contributions from recent infrequent bursts such that \cite{Cui:2019kkd}
\begin{align}
\int_0^{z_{\rm min}} dz \, \Theta(t_i-t_F) \int dl \,\frac{d^2R}{dz dl} = f .
\end{align}
The waveform is
\begin{align}
h(f,l,z) = g_{1c}\frac{G\mu \, l^{2/3}}{(1+z)^{1/3}r(z)}f^{-4/3},
\end{align}
with $g_{1c}\simeq 0.85$ \cite{LIGOScientific:2021nrg}. We have taken the integration limit on $l$ to be from $0$ to $2t$ ($3t$) for the radiation (matter) domination. However, the $\Theta$ functions will control the upper and lower limits during numerical evaluations.

{Following Appendix A of Refs.~\cite{Siemens:2006vk,LIGOScientific:2017ikf}, the Hubble parameter in $\Lambda${CDM} is expressed as
\begin{align}
H(z)=H_0\mathcal{H}(z),
\end{align}
where
\begin{align}\label{eq:mathcalH}
\mathcal{H}(z)=\sqrt{\Omega_{\Lambda,0}+\Omega_{m,0}(1+z)^3+\Omega_{r,0}\mathcal{G}(z)(1+z)^4} \ .
\end{align}
Here $\Omega_{i,0}$ is the fractional energy density in the $i$th component, and the subscripts $m$, 
$r$, and $\Lambda$ stand for matter, radiation, and the cosmological constant, respectively.
We take $\Omega_{m,0}=0.308$, $\Omega_{r,0}=9.1476\times 10^{-5}$, $\Omega_{\Lambda,0}=1-\Omega_{m,0}-\Omega_{r,0}$. The quantity $\mathcal{G}(z)$ in Eq.~(\ref{eq:mathcalH}) is given by \cite{Binetruy:2012ze}
\begin{align}
\mathcal{G}(z)=\frac{g_*(z)g_{*S}^{4/3}(0)}{g_*(0)g_{*S}^{4/3}(z)} \, ,
\end{align}
with $g_*$ and $g_{*S}$ being the effective numbers of relativistic degrees of freedom for the energy and entropy densities respectively. We set \cite{LIGOScientific:2017ikf}
\begin{align}
\mathcal{G}(z)=\begin{cases}
1 \ \ &\mathrm{for} \ z<10^9 , \\
0.83 \ \ &\mathrm{for} \ 10^9<z<2\times 10^{12} , \\
0.39 \ \ &\mathrm{for} \ z>2\times 10^{12} .
\end{cases}
\end{align}}
We can express the cosmic time $t$, proper distance $r$, and proper differential volume element $dV$as functions of the cosmological redshift:
\begin{align}\label{eq:t}
t(z)=H_0^{-1}\phi_t(z) \ \ \mathrm{with} \  \ \phi_t(z) = \int_z^\infty\frac{dz'}{\mathcal{H}(z')(1+z')} \ ,
\end{align} 
\begin{align}\label{eq:r}
r(z)=H_0^{-1}\phi_r(z) \ \ \mathrm{with} \  \ \phi_r(z) = \int_0^z \frac{dz'}{\mathcal{H}(z')} \ ,
\end{align}
and
\begin{align}
dV(z)=H_0^{-3}\phi_V(z) dz \ \ \mathrm{with} \  \ \phi_V(z) = \frac{4\pi \phi_r^2(z)}{(1+z)^3\mathcal{H}(z)} \ .
\end{align}

\subsection{Gravitational Waves from Monopole-antimopole Pairs Connected by Strings}
In this section we compute the gravitational wave spectrum from the $MS\bar{M}$ structures following Refs.~\cite{Martin:1996ea,Martin:1996cp,Leblond:2009fq}, assuming that the monopoles do not carry unconfined fluxes.
The number density of the monopole-antimonopole pairs at any time $t(z)$ after their horizon reentrance at $t_M(z_M)$ during radiation dominance is given by
\begin{align}
\label{eq:density_mm_pair}
\tilde{n}(z)=(2t_M)^{-3}\left(\frac{1+z}{1+z_M}\right)^3 .
\end{align}
The length of the $MS\bar{M}$ structure with monopole mass $m_M$ and string tension $\mu$ is given by
\begin{align}
\tilde{l}(z) = 2t_M - \tilde{\Gamma}G\mu(t(z)-t_M), 
\end{align}
and the maximum Lorentz factor for the monopoles is
\begin{align}
\gamma \sim \frac{\mu}{m_M}\tilde{l} \ . 
\end{align}
We have taken $\tilde{\Gamma}\sim 8\ln \gamma(z_M)$ with the initial size of each $MS\bar{M}$ structure being equal to the particle horizon $2t_M$. The $MS\bar{M}$ structures oscillate with a period $T\sim \tilde{l}$ and the number of gravitational wave bursts per unit space time volume at time $t$ is
\begin{align}
\nu(z)\sim \frac{\tilde{n}(z)}{\tilde{l}(z)} \, .
\end{align}
The fraction of the bursts that can be observed is \cite{Leblond:2009fq}
\begin{align}
\Delta(f,z)\sim \frac{1}{4(1+z)f\tilde{l}}\Theta(\gamma-1)\Theta((1+z)f\tilde{l}-1)\Theta(\gamma^2-(1+z)f\tilde{l}) \ .
\end{align}

For $\gamma\gg 1$, most of the gravitational radiation is radiated within the frequency range \cite{Martin:1996ea,Martin:1996cp,Leblond:2009fq}
\begin{align}
\frac{1}{(1+z)\tilde{l}}<f<\frac{\gamma^2}{(1+z)\tilde{l}} \, ,
\end{align}
and the wave form, i.e. the amplitude of the Fourier transform of the trace of the metric perturbation for each burst, is given by \cite{Martin:1996ea,Martin:1996cp,Leblond:2009fq}
\begin{align}
h(f,z)=\tilde{g}\frac{G\mu \tilde{l}(z)}{r(z)}f^{-1}  ,
\end{align}
where $\tilde{g}=2\sqrt{2}\sin^2\sqrt{2}/\pi$ and $r$ is defined in Eq.~(\ref{eq:r}). 

The burst rate of gravitational waves observed at frequency $f$ from a volume $dV(z)$ at redshift $z$ is given by
\begin{align}\label{eq:dv}
\frac{dR}{dV(z)} = (1+z)^{-1}\nu(z)\Delta(f,z) \, ,
\end{align}
where $(1+z)^{-1}$ is the redshift factor. The observed burst rate at redshift $z$ is therefore
\begin{align}
\frac{dR}{dz} = H_0^{-3}\phi_V(z) (1+z)^{-1}\nu(z)\Delta(f,z) \ .
\end{align}
The stochastic gravitational wave background from the $MS\bar{M}$ structures will then be
\begin{align}\label{eq:GWs-Omega-MSM}
\Omega_{GW}^{MS\bar{M}}(f) = \frac{4\pi^2}{3H_0^2}f^3\int_{z_*}^{z_M}dz \, h^2(f,z)\frac{dR}{dz} \ .
\end{align}
Here the lower limit $z_*$ in the integral is to deduct the contributions from the infrequent burst events and is found from
\begin{align}
\int_0^{z_*}dz \,\frac{dR}{dz} = f.
\end{align}

Finally, the total gravitational radiation from the quasi-stable strings is equal to the sum of Eqs.~(\ref{eq:GWs-Omega-cusps}) and (\ref{eq:GWs-Omega-MSM}), 
\begin{align}
\Omega_{GW}(f) = \Omega_{GW}^{\rm loop}(f) + \Omega_{GW}^{MS\bar{M}}(f) \ .
\end{align}

\section{Results}
\label{sec:results} 

For definiteness, we apply our calculation of the gravitational wave spectra to a class of non-supersymmetric GUT models \cite{Chakrabortty:2020otp} with two intermediate scales $M_I$ and $M_{II}$ and inflation driven by the Coleman-Weinberg potential of a gauge singlet real scalar field 
\cite{Shafi:1983bd,Lazarides:1984pq}. The higher scale $M_I$ is associated with the production of magnetic monopoles, and 
$M_{II}$ is linked to the generation of a network of quasi-stable cosmic strings connecting the monopoles. We first assume that both the monopoles and stings are generated during inflation, undergo partial inflation, and eventually reenter the post-inflationary horizon at a cosmic time 
$t_{\rm def}$ equal to $t_M$ or $t_F$, respectively. Of course, the monopoles are generated prior to the strings and, therefore, experience more $e$-foldings. Consequently, they reenter the horizon after the strings.

In order to calculate the horizon reentry time $t_{\rm def}$ of topological defects (monopoles or strings), we follow the analysis of Ref.~\cite{Lazarides:2021uxv} (see also Ref.~\cite{Chakrabortty:2020otp}), which we now summarize. The phase transition during which the intermediate gauge symmetry breaking is completed and the topological defects are formed occurs when the Ginzburg criterion \cite{ginzburg}
\begin{equation}
\frac{4\pi}{3}H(\phi)^{-3}\Delta V=T_H
\label{criterion}
\end{equation}
is satisfied. Here $\Delta V=(m^{\rm def}_{\mathrm{eff}})^{4}/16\alpha$, where $m_{\rm eff}^{\rm def}$ is the effective mass at the phase transition of the normalized real scalar field $\chi$ causing the symmetry breaking and is given by
\begin{equation}
\label{effective-mass_chi1}
({m^{\rm def}_{\mathrm{eff}}})^{2}=2[\beta^2\phi^2-\sigma T_H^2] \ ,
\end{equation}
with $\beta=\sqrt{\alpha}M_{\rm def}/M$. Also $\alpha\simeq 0.25$ is the quartic coupling constant of $\chi$, $M_{\rm def}$ ($=M_I$ or $M_{II}$) is the defect scale, and $M$ denotes the inflaton vacuum expectation value. We set $\sigma\simeq 1$, $T_H=H(\phi)/2\pi$ is the Hawking temperature with $H(\phi)$ being the Hubble parameter at the transition, and $\phi$ is the corresponding value of the inflaton field, which can be estimated from
\begin{equation}\label{breaking_scale_2}
M_{\rm def} = \sqrt{\left(2\sqrt{6}\pi\alpha^{\frac{1}{2}}+\sigma\right)}\frac{H(\phi)}{2\pi\phi}\frac{M}{\sqrt{\alpha}}.
\end{equation} 

\begin{figure}[htbp]
\centering
\includegraphics[width=0.7\textwidth]{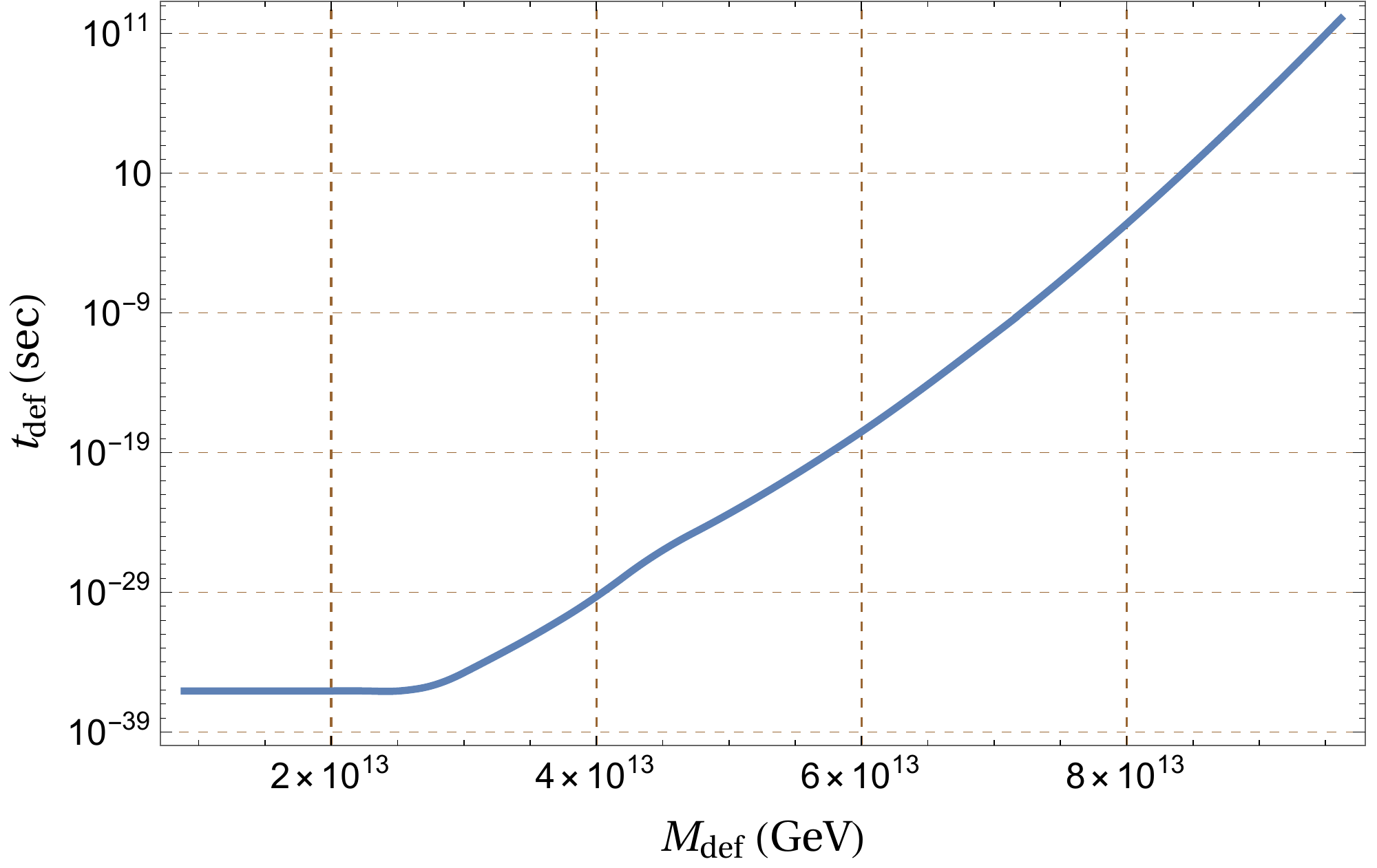}
\caption{Horizon reentry time $t_{\rm def}$ of the topological defects (monopoles or strings) versus the symmetry breaking scale $M_{\rm def}$ ($M_I$ or $M_{II}$).}
\label{reentry}
\end{figure}
\begin{figure}[htbp]
\centering
\subfloat[][]{\includegraphics[width=0.47\textwidth]{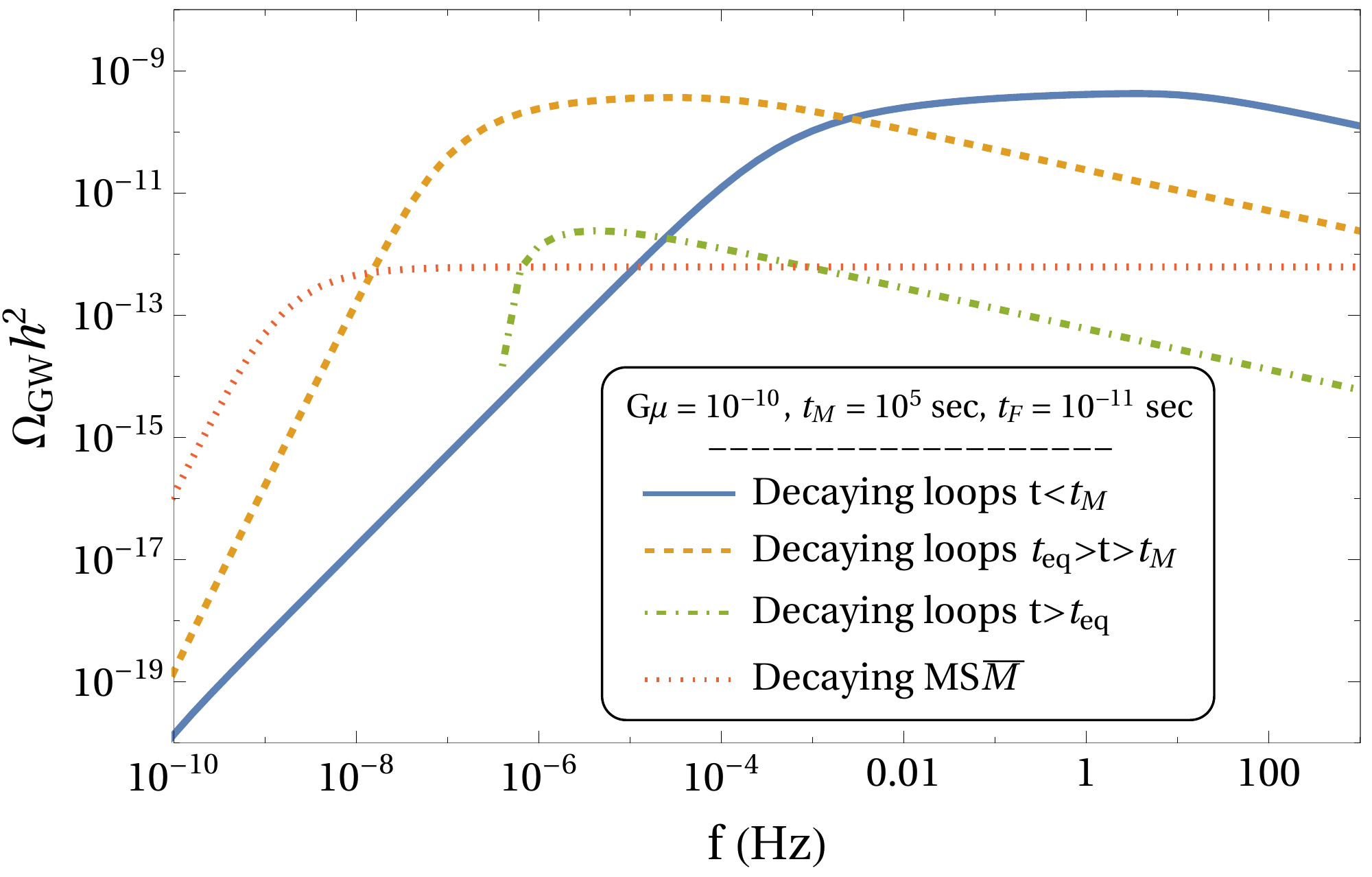}}
\hspace{0.05\textwidth}
\subfloat[][]{\includegraphics[width=0.47\textwidth]{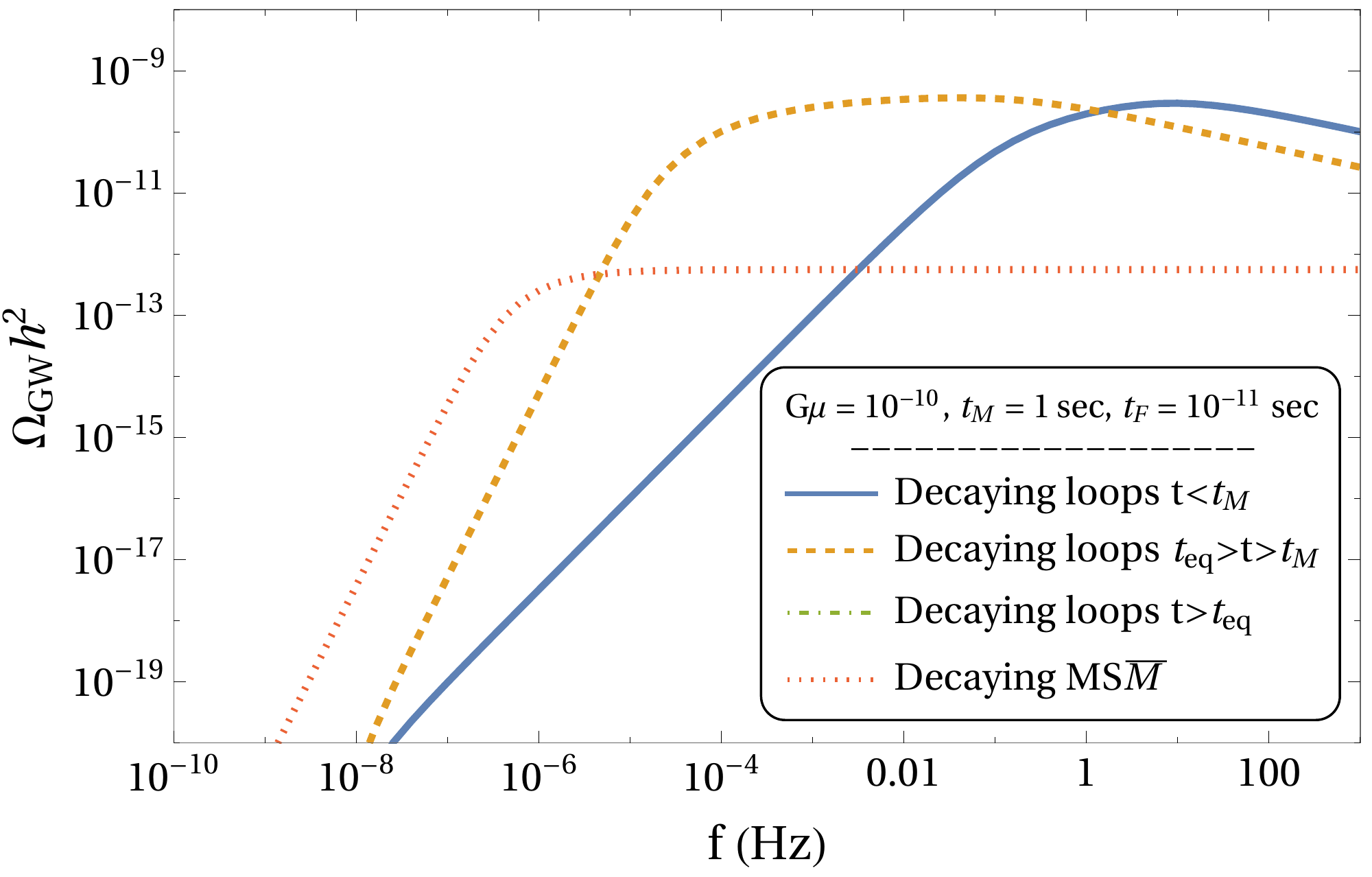}} 
\caption{The gravitational wave spectra from loops decaying before and after $t_M$ (the horizon reentrance time of the monopoles) during radiation dominance, from loops decaying after the equidensity time $t_{eq}$, and from the decaying $MS\bar{M}$ structures. The plots are for 
$G\mu=10^{-10}$ and thus $t_F\simeq 10^{-11}$ sec. In panel (a) $t_M=10^{5}$ sec and the
monopole mass $m_M=4.45\times 10^{15}$ GeV, and in panel (b) $t_M=1$ sec and $m_M=4.17\times 
10^{15}$ GeV.}
\label{fig:GWs_components}
\end{figure}

At production the mean inter-defect distance is $p \, H(\phi)^{-1}$, where $p\simeq 2$ is a geometric factor. This distance is subsequently stretched by the factors $\exp\left(N_{\rm def}\right)$, $\left(t_r/\tau\right)^{2/3}$, and $\left(t/t_r\right)^{1/2}$, respectively, during inflation, inflaton oscillations, and after reheating until time $t$ in the radiation dominated Universe. Here, $N_{\rm def}$ is the number of $e$-foldings that the defects underwent, $t_r\simeq 
2.37\times 10^{-25}~{\rm sec}$ is the reheat time, and $\tau\simeq 8.3\times 10^{-37}~{\rm sec}$ is the cosmic time at the end of inflation. If the cosmic time $t$ lies in the era of inflaton oscillations, the only stretching factor of the inter-defect distance after the end of inflation is $\left(t/\tau\right)^{2/3}$. Equating the resulting inter-defect distance with the particle horizon, we find the horizon reentrance time $t_{\rm def}$ of the defects. 
In Fig.~\ref{reentry} we depict the horizon reentry time $t_{\rm def}$ of the defects (monopoles or strings) versus the symmetry breaking scale $M_{\rm def}$. We observe that the curve is terminated on the left at the time $\tau$ corresponding to the end of inflation.  

We take $t_M=\lbrace 10^{-5}, 1, 10^{5}, 10^{10}\rbrace$ sec, which, as deduced from Fig.~\ref{reentry}, correspond to $M_{I}=\lbrace 7.7\times 10^{13}, 8.3\times 10^{13}, 8.86\times 10^{13}, 9.4\times 10^{13}\rbrace$ GeV, respectively, with the gauge coupling $g\simeq 0.5$. These scales yield monopole masses $m_M=\lbrace 3.87\times 10^{15}, 4.17\times 10^{15}, 4.45\times 10^{15}, 4.72\times 10^{15}\rbrace$ GeV, respectively.

We consider strings with dimensionless tension $G\mu=\lbrace 10^{-10}, 10^{-11}\rbrace$
which yield $M_{II}=\lbrace 6.87\times 10^{13}, 2.17\times 10^{13}\rbrace$ GeV. From Fig.~\ref{reentry}, we observe that in both these cases, the strings are generated during inflation and reenter the post-inflationary horizon at $t_F\simeq \lbrace 10^{-11}, 8.3\times 10^{-37}\rbrace$ sec, respectively. We note that, for $G\mu=10^{-11}$ the horizon reentry occurs during the inflaton oscillations, very close to the end of inflation, and well before the reheat time 
$\sim 10^{-25}$ sec. We also consider strings with $G\mu=\lbrace 10^{-12}, 10^{-13}, 10^{-14}, 10^{-15}\rbrace$, which are generated for the first time after the end of inflation and during inflaton oscillations. We have checked that, for $G\mu\leq 10^{-11}$, the contribution to the gravitational wave spectrum of the loops generated during inflaton oscillations is quite negligible and, therefore, we set $t_F=10^{-25}$ sec in Eq.~(\ref{eq:GWs-Omega-cusps}). All scales $M_{II}$ are smaller than the scales $M_{I}$ and the quasi-stability condition for the strings is very well satisfied in all cases.
\begin{figure}[htbp]
\centering
\subfloat[][]{\includegraphics[width=0.47\textwidth]{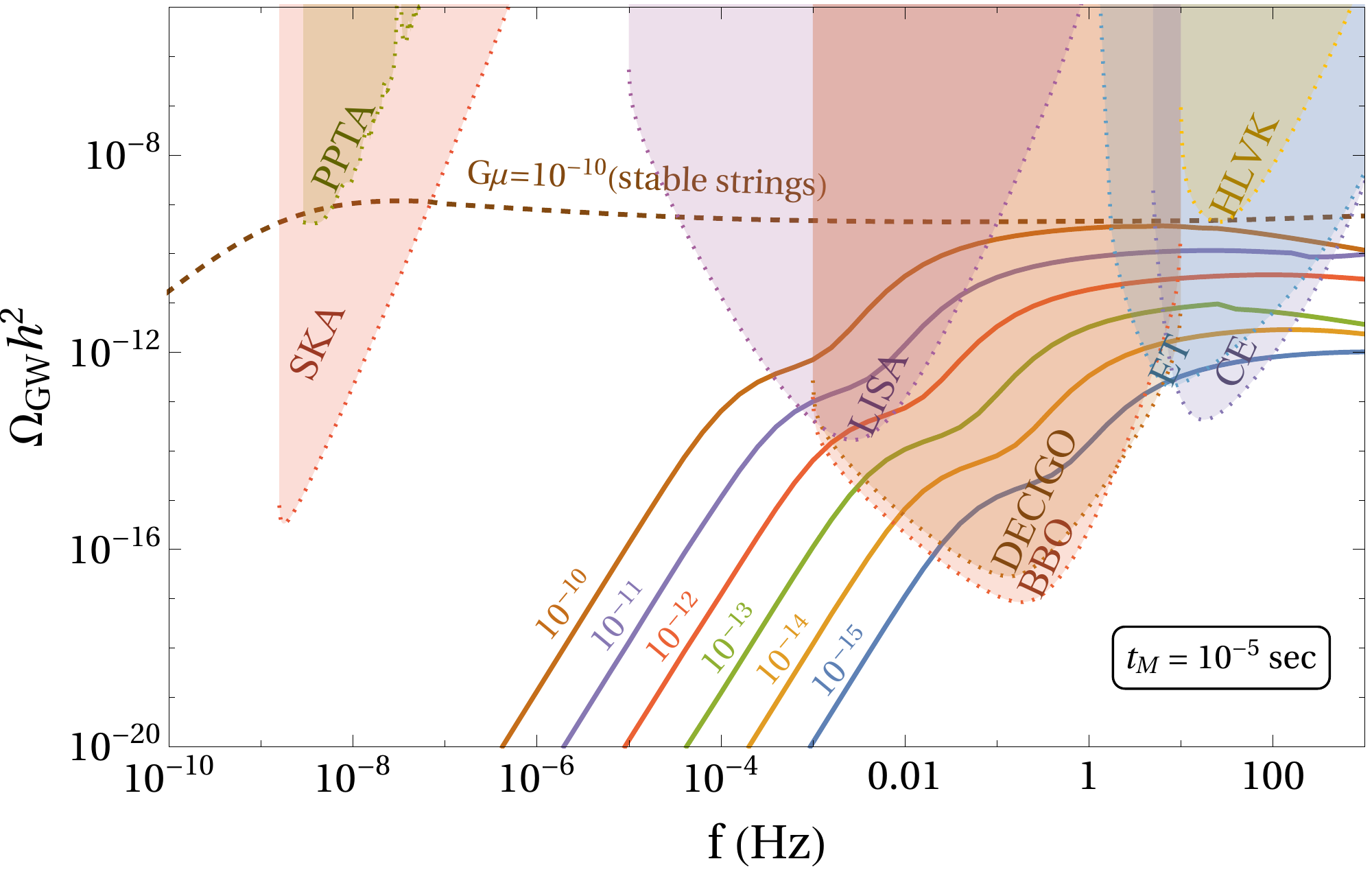}\label{subfig:GWs_meta_tM_10m5_tF_CW}} 
\hspace{0.05\textwidth}
\subfloat[][]{\includegraphics[width=0.47\textwidth]{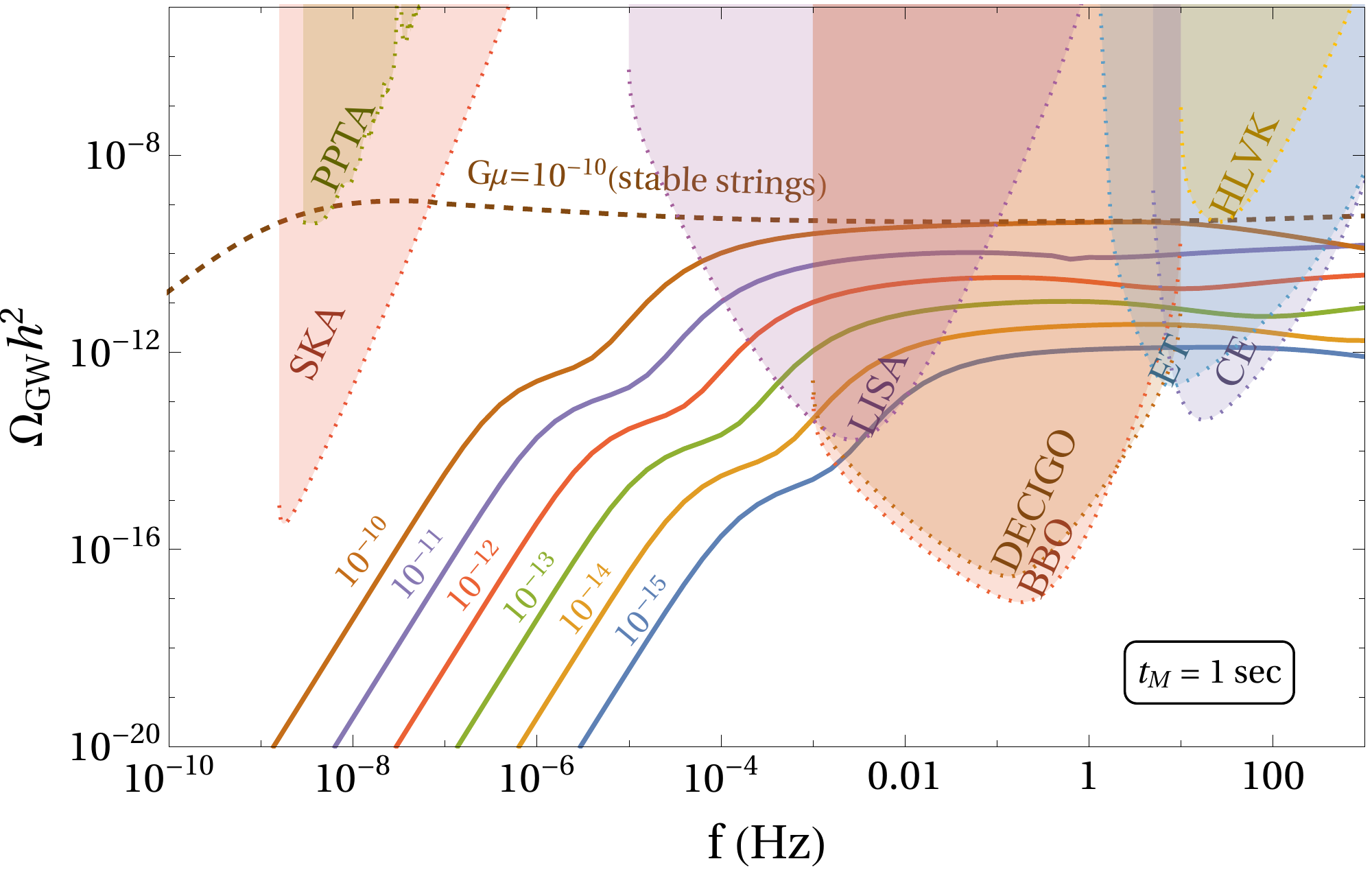}\label{subfig:GWs_meta_tM_10p0_tF_CW}} 
 \\
\subfloat[][]{\includegraphics[width=0.47\textwidth]{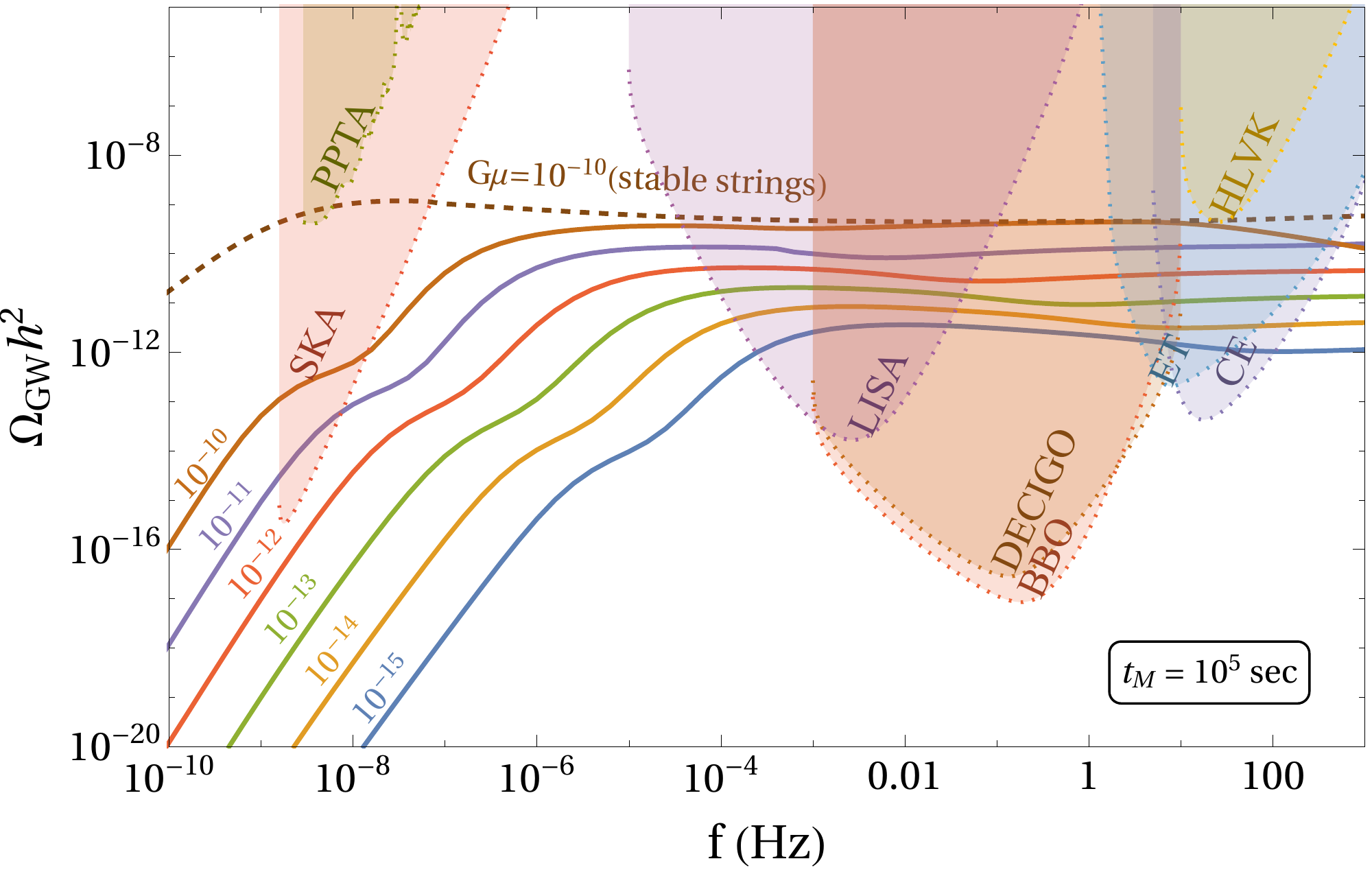}\label{subfig:GWs_meta_tM_10p5_tF_CW}} 
\hspace{0.05\textwidth}
\subfloat[][]{\includegraphics[width=0.47\textwidth]{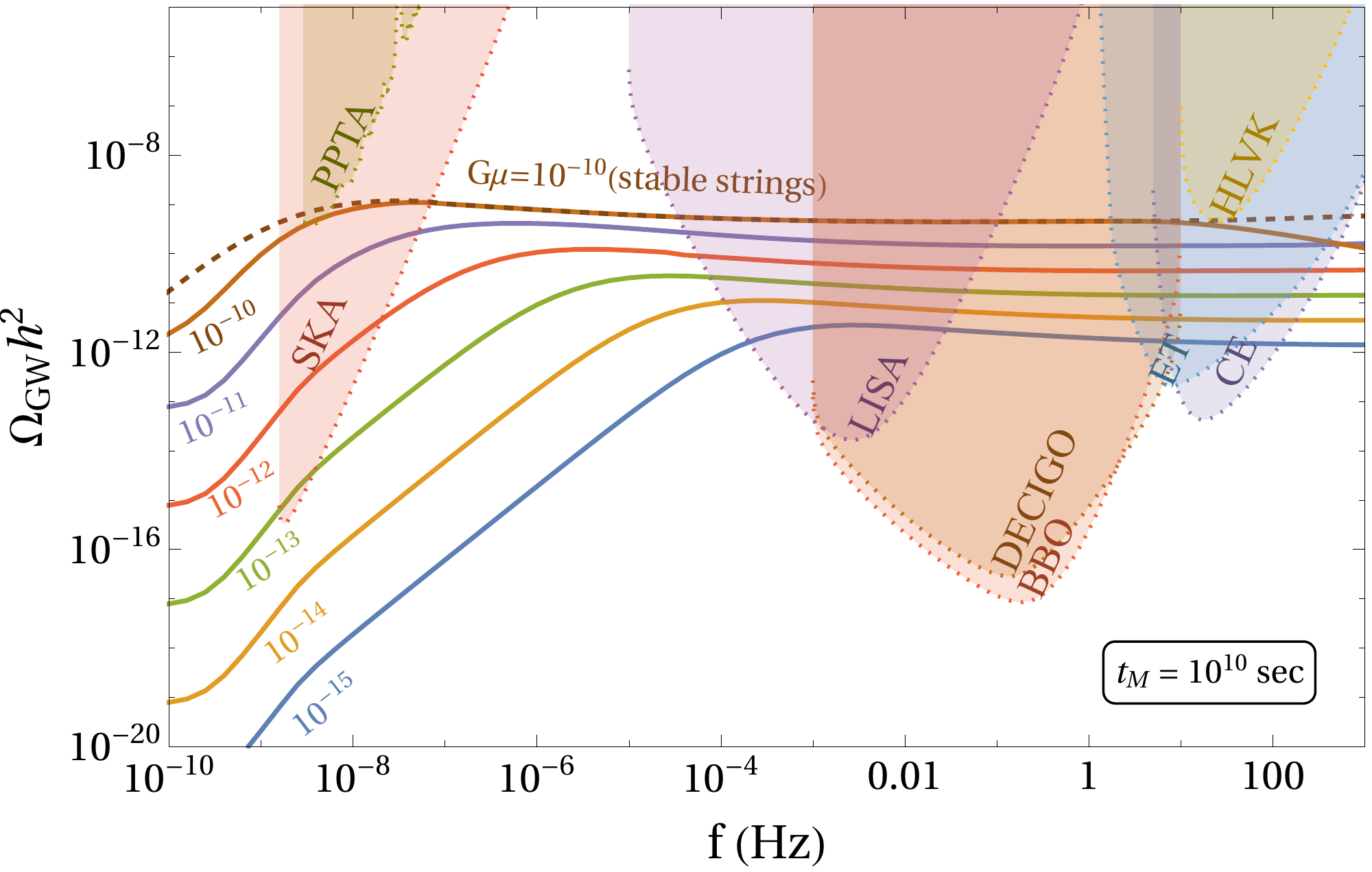}\label{subfig:GWs_meta_tM_10p10_tF_CW}}
 \caption{The total gravitational wave spectra from quasi-stable cosmic strings with varying 
$G\mu$ values as indicated and for four different horizon reentrance times of the monopole-antimonopole pairs $t_M=\lbrace 10^{-5}, 1, 10^5, 10^{10}\rbrace$ sec. The horizon reentry time of the strings $t_F$ is compatible with inflation driven by the Coleman-Weinberg potential of a real GUT singlet for each choice of $G\mu$, namely we take $t_F=10^{-11}$ sec for $G\mu=10^{-10}$ and 
$t_F=10^{-25}$ sec in all other cases. We include, for comparison, the gravity wave spectrum from a  network of stable strings with $G\mu=10^{-10}$ that experience no inflation (brown dashed line). The sensitivity curves \cite{Thrane:2013oya, Schmitz:2020syl} for PPTA \cite{Shannon:2015ect} and various proposed experiments, namely, SKA \cite{5136190, Janssen:2014dka}, CE \cite{PhysRevLett.118.151105}, ET \cite{Mentasti:2020yyd}, LISA \cite{Bartolo:2016ami, amaroseoane2017laser}, DECIGO \cite{Sato_2017}, BBO \cite{Crowder:2005nr, Corbin:2005ny}, HLVK \cite{KAGRA:2013rdx}, etc. are shown on the plots.}\label{fig:GWs_meta_strng}
\end{figure}

Fig.~\ref{fig:GWs_components} shows the gravitational wave spectra from the string loops decaying before and after $t_M$ during radiation dominance, from the loops decaying after the equidensity time $t_{eq}$, and from the $MS\bar{M}$ structures. The plots are for $G\mu=10^{-10}$ and thus $t_F\simeq 10^{-11}$ sec. In panel (a) $t_M=10^{5}$ sec, $M_I=8.86\times 10^{13}$ GeV, and the monopole mass $m_M=4.45\times 10^{15}$ GeV. In panel (b) $t_M=1$ sec, $M_I=8.3\times 10^{13}$ GeV, and 
$m_M=4.17\times 10^{15}$ GeV. We note that, as $t_M$ decreases, the contributions to the gravitational wave spectrum from the loops decaying before $t_{eq}$ and the $MS\bar{M}$ structures are drastically reduced at lower frequencies, but at higher frequencies they are very little affected. The contribution from the loops decaying after $t_{eq}$, on the contrary, is drastically reduced at all frequencies as $t_M$ decreases.    

At this point, we could compare the gravitational wave spectra from $MS\bar{M}$ segments with those computed using the normal mode methods as in Refs.~\cite{Buchmuller:2021mbb,Dunsky:2021tih}. In Ref.~\cite{Dunsky:2021tih}, the $MS\bar{M}$ structures exist within the particle horizon from a very early time. They contribute to the high frequency regime of the gravitational wave spectrum because their size is very small. However, in our case, the monopoles reenter the horizon at a comparatively very late time and we get gravitational waves starting from much lower frequencies. Also, there are some dissipation effects at a very early time which were taken into account in Ref.~\cite{Dunsky:2021tih}. Therefore, we cannot compare this result with ours. We may compare our result with the gravitational waves from segments\footnote{The segments originate from `metastable' strings.} obtained in Ref.~\cite{Buchmuller:2021mbb}. From Eqs.~(37) and (46) of this reference, we find that the height of the plateau is of the order of $\Gamma^{-1/2}$ and $\tilde{\Gamma}^{-3/2}$ in the case of loops and segments respectively. As a consequence, we expect that there is an $\mathcal{O}(100)$ suppression of the gravitational waves from the $MS\bar{M}$ structures compared to that from the loops which agrees quite well with our Fig.~\ref{fig:GWs_components}.

Fig.~\ref{fig:GWs_meta_strng} shows the gravitational wave spectra from quasi-stable cosmic strings with different $G\mu$ values and for four different horizon reentry times of the monopole-antimonopole pairs, namely, $t_M=\lbrace 10^{-5}, 1, 10^5, 10^{10}\rbrace$ sec. The horizon reentry time 
$t_F$ of the strings is compatible with Coleman-Weinberg inflation for each choice of $G\mu$. We find that for $G\mu = 10^{-10}$, the strings reenter the horizon around $t_F\sim 10^{-11}$ sec during radiation dominance. For $G\mu=10^{-11}$ or lower, the strings either reenter the horizon or are produced for the first time during inflaton oscillations. However, the contribution to the spectrum from loops generated until reheating is quite negligible in the frequency range under consideration. Therefore, in these cases, we set $t_F=10^{-25}$ sec, which lies close to the reheating time. For comparison, we include the spectrum for a network of stable strings with $G\mu=10^{-10}$ without inflation, which violates the PPTA bound. We observe that at intermediate and especially at lower frequencies, the spectrum for quasi-stable strings is reduced relative to the spectrum of stable strings with the reduction being more pronounced as $t_M$ decreases. On the other hand, at higher frequencies, the reduction is limited and the spectrum is practically insensitive to the value of $t_M$. In summary, we find that in the quasi-stable string scenario, values of $G\mu$ up to about $10^{-10}$, which are accessible in the ongoing and future experiments, are acceptable without coming into conflict with the PPTA bound.    
\section{GUTs with Quasi-stable Cosmic Strings}
For completeness let us briefly outline two realistic models in which the quasi-stable cosmic string scenario can be realized. Our first example is based on a non-supersymmetric $SO(10)$ model which breaks to the SM gauge group as follows:
\begin{align*}
SO(10)\xrightarrow{M_{\rm GUT}} SU(4)_c\times &SU(2)_L\times SU(2)_R  \xrightarrow{M_{I}} SU(3)_c\times U(1)_{B-L}\times SU(2)_L\times U(1)_R \\ & \xrightarrow{M_{II}} SU(3)_c\times SU(2)_L\times U(1)_Y .
\end{align*}

The first breaking yields the superheavy GUT monopole \cite{tHooft:1974kcl,Lazarides:1980cc}, which we assume is inflated away.
With two $U(1)$’ s appearing from the $SU(4)_c\times SU(2)_L\times SU(2)_R$ breaking at the first intermediate scale $M_I$ to $SU(3)_c\times U(1)_{B-L}\times SU(2)_L\times U(1)_R$, we obtain two sets of monopoles, referred to here, following Ref.~\cite{Lazarides:2019xai}, as red and blue monopoles. The next breaking at a lower intermediate scale $M_{II}$ yields, among other things, the quasi-stable cosmic strings, which are associated with the broken $U(1)$ generator orthogonal to $U(1)_Y$.
After partial inflation the red and blue monopoles enter the horizon and some get attached to their corresponding antimonopoles via quasi-stable strings. The red and blue monopoles can also merge to produce stable doubly charged electromagnetic monopoles that also carry color magnetic field, and they may be present in our galaxy at a potentially observable level \cite{Lazarides:1984pq,Senoguz:2015lba,Chakrabortty:2020otp}.
Note that if $U(1)_{B-L}$ is broken with a $126$-plet Higgs we also produce topologically stable intermediate scale strings \cite{Kibble:1982ae}, and these also will contribute to the gravitational wave spectrum. For other $SO(10)$ symmetry breaking patterns, see Refs.~\cite{Chakrabortty:2009xm,Chakrabortty:2017mgi,Chakrabortty:2019fov,King:2020hyd,King:2021gmj,Holman:1982tb}.

A second realistic example of quasi-stable strings is provided by the trinification gauge symmetry 
$SU(3)_c\times SU(3)_L\times SU(3)_R$. Following Ref.~\cite{Lazarides:2021tua}, we first break
$SU(3)_L\times SU(3)_R$ to $SU(2)_L \times U(1)_L \times SU(2)_R \times U(1)_R$ at some scale $M_I$ which can be smaller than $M_{\rm GUT}$ because of the absence of gauge bosons that mediate proton decay. Analogous to the $SO(10)$ case, we obtain two sets of monopoles which can be partially inflated. The next breaking to the SM gauge symmetry at a lower intermediate scale $M_{II}$ yields a topologically stable magnetic monopole carrying three quanta of Dirac magnetic charge \cite{Lazarides:2021tua}. In addition we obtain the quasi-stable cosmic string network we are after.
Indeed these two realistic examples yield additional topological structures such as necklaces \cite{Lazarides:2019xai} that we plan to explore elsewhere.

 It should be mentioned that in these examples the monopoles carry unconfined fluxes too. This leads to the decay of the $MS\bar{M}$ structures predominantly by the emission of massless gauge bosons. Consequently, their contribution to the gravitational wave spectrum is suppressed. However, this does not affect our overall results in any drastic way since, as one can see from Fig.~\ref{fig:GWs_components}, this contribution is any way subdominant for most frequencies. Moreover, one can show that the radiation into massless gauge bosons has no measurable effect on the evolution of the Universe.       
\section{Conclusions}
We have estimated the stochastic gravitational wave spectrum emitted by what we have termed `quasi-stable' cosmic strings. Intermediate scale magnetic monopoles are created prior to the cosmic strings and experience partial inflation. The strings being an order of magnitude or more lighter than the monopoles are quantum mechanically stable. They also may experience partial inflation and enter the post-inflationary horizon later or be created after the end of inflation for the first time. In any case, as the strings are within the horizon, they form random walks with step of the order of the horizon terminating on the monopoles which are still out of the horizon. They inter-commute generating loops which decay into gravitational waves. However, as the monopoles reenter the horizon we obtain monopole-antimonopole pairs connected by string segments which also decay into gravitational waves. This effectively results in the absence of long string loops and segments, thereby leading to a reduction of the gravity wave spectrum in the low frequency region. Consequently, the spectrum is accessible in the ongoing and future experiments without being in conflict with the PPTA bound. We provide examples of the gravitational wave spectrum for $G \mu$ values varying between $10^{-10}$ and $10^{-15}$. We present two examples of GUTs that predict the presence of these quasi-stable strings accompanied by the somewhat heavier monopoles.

\section{Acknowledgments}
This work is supported by the Hellenic Foundation for
Research and Innovation (H.F.R.I.) under the “First Call
for H.F.R.I. Research Projects to support Faculty Members and Researchers and the procurement of high-cost
research equipment grant” (Project Number: 2251).
\appendix
\section*{Appendix: Estimation of Loop Distributions}\label{appen:loop_dist}
\addcontentsline{toc}{section}{Appendix: Estimation of Loop Distributions}

To estimate the loop number density per unit length at time $t$ we follow Ref.~\cite{Blanco-Pillado:2013qja}. We define the loop size parameter $x$ of a string loop of length $l$ at time $t$ as
\begin{align}
x = \frac{l(t)}{d_h(t)} \, ,
\end{align}
where the particle horizon $d_h = 2t$ $(3t)$ during the radiation (matter) dominated Universe. The number of loops within the horizon volume $d_h^3$ per unit $x$ during the radiation domination is given by \cite{Blanco-Pillado:2013qja}
\begin{align}\label{eq:n_x_rad}
n_r(x) = \frac{0.52 \,\Theta(0.05 - x)}{(x+\Gamma G\mu/2)^{5/2}} \, .
\end{align}

We assume that after the horizon reentrance of the monopole-antimonopole pairs, the loop formation ceases and there exist only the decaying loops which are remnants of those formed at or before 
$t_M$. A loop of length $l$ at time $t>t_M$ had a length at $t_M$ given by
\begin{align}\label{eq:l_x_t_gtr_tM}
l_M \equiv  2x_Mt_M = l(t) +\Gamma G\mu(t-t_M)\, .
\end{align} 
The length of a loop in the radiation (or matter) dominated era is $l=2xt$ (or $3xt$). There are 
$n_r(x_M)dx_M$ string loops in the volume $(2t_M)^3$ at time $t_M$ within the range $dx_M$ of the loop size parameter around $x_M$, and thus $n_r(x_M)dx_M (a(t_M)/2t_M)^3$ is the number of loops per comoving volume. Equating this with the number of loops in the same comoving volume in the range 
$t_{eq}>t>t_M$ we obtain
\begin{align}\label{eq:n_x_rad_t_gtr_tM}
n_r(x;t_{eq}>t>t_M) = n_r(x_M)\frac{\partial x_M}{\partial x} \left(\frac{a(t_M)}{a(t)}\right)^3\left(\frac{2t}{2t_M}\right)^3 \, .
\end{align}

The scale factor $a(t)\propto t^{1/2}$ in the radiation dominated Universe. Substituting the expression in Eq.~(\ref{eq:n_x_rad}) for $n_r(x_M)$ and ${\partial x_M}/{\partial x}$ from Eq.~(\ref{eq:l_x_t_gtr_tM}) into Eq.~(\ref{eq:n_x_rad_t_gtr_tM}) we find
\begin{align}\label{eq:n_x_rad_t_gtr_tM_2}
n_r(x;t>t_M) = \frac{0.52 \,\Theta(0.05-x_M)}{(x_M+\Gamma G\mu/2)^{5/2}}\left(\frac{t}{t_M}\right)^{5/2} .
\end{align}
Finally, the number density of loops of size $l$ and time $t$ is given by
\begin{align}\label{eq:n_l_rad_t_gtr_tM} 
n_r(l,t>t_M) = \frac{n_r(x;t>t_M)}{d_h(t)^4} \, ,
\end{align}
and after substituting $x_M$ from Eq.~(\ref{eq:l_x_t_gtr_tM}) into 
Eq.~(\ref{eq:n_x_rad_t_gtr_tM_2}), and using Eq.~(\ref{eq:n_l_rad_t_gtr_tM}) we obtain Eq.~(\ref{eq:n-loop-rad_after_tM}).

Similarly, the number of loops per unit loop size parameter at $x$ within the horizon volume $d_h^3$ during the matter dominated era is 
\begin{align}\label{eq:n_x_mat_t_gtr_tM}
n_r(x;t>t_M) = n_r(x_M)\frac{\partial x_M}{\partial x} \left(\frac{a(t_M)}{a(t)}\right)^3\left(\frac{3t}{2t_M}\right)^3.
\end{align}
The number of loops surviving in the matter dominated era is obtained from Eq.~(\ref{eq:n_x_mat_t_gtr_tM}) and Eq.~(\ref{eq:l_x_t_gtr_tM}), namely
\begin{align}\label{eq:n_x_mat_t_gtr_tM_2}
n_{rm}(x;t>t_{eq}) =\frac{81}{16} \frac{0.52 \,\Theta(0.05-x_M)}{(x_M+\Gamma G\mu/2)^{5/2}}\left(\frac{t}{t_M}\right)^{5/2}\frac{t_{eq}^{1/2}}{t_M^{5/2}}t^2 \, .
\end{align}
In physical units we can express the number density of loops as $n_{rm}(l,t>t_{eq}) \equiv n_{rm}(x;t>t_{eq})/(3t)^4$, which leads to Eq.~(\ref{eq:n-loop-radmat}).
\bibliographystyle{mystyle}
\bibliography{strng}

\end{document}